\documentclass[twocolumn,preprintnumbers,superscriptaddress]{revtex4}

\usepackage{amsmath}
\usepackage{graphics}
\usepackage{graphicx}
\usepackage[mathcal]{eucal}
\usepackage{color}

\begin{document}

\preprint{UCI-TR-2009-20}

\title{TeV Scale Seesaw and a flavorful $Z^{\prime}$ at the LHC}

\author{Mu-Chun Chen}
\email[]{muchunc@uci.edu}
\affiliation{Department of Physics \& Astronomy, 
University of California, Irvine, CA 92697-4575, USA}
\author{Jinrui Huang}
\email[]{jinruih@uci.edu}
\affiliation{Department of Physics \& Astronomy, 
University of California, Irvine, CA 92697-4575, USA}

\date{\today}

\begin{abstract}
\label{abstract}
Small neutrino masses and their large mixing angles can be generated at the TeV scale by augmenting the Standard Model with an additional generation dependent, anomaly-free $U(1)_{\nu}$ symmetry, in the presence of three right-handed neutrinos. The $Z^{\prime}$ gauge boson associated with the breaking of the $U(1)_{\nu}$ symmetry can be produced at the LHC. The flavorful nature of the $Z^{\prime}$ can be established by measuring its non-universal couplings to the charged leptons as determined by the lepton's $U(1)_{\nu}$ charges, which also govern the neutrino flavor structure. While the LHC has the potential of discovering the $Z^{\prime}$ up to $M_{Z^{\prime}} = 4.5$ TeV with 100 fb$^{-1}$ data at the center of mass energy $\sqrt{s} = 14$ TeV, to establish the flavorful nature of the $Z^{\prime}$ requires much higher integrated luminosity. For our bench mark parameters that are consistent with neutrino oscillation data, at $\sqrt{s} = 14$ TeV, a $5\sigma$ distinction between the dielectron and dimuon channels for $M_{Z^{\prime}} = 3$ TeV requires 500 fb$^{-1}$ of data. We find that the forward backward asymmetry distributions can also be useful in distinguishing the dielectron and dimuon channels in the low invariant mass and transverse momentum regions. 
\end{abstract}


\maketitle

\section{Introduction}
\label{sec:intro}

The $Z^{\prime}$ gauge boson~\cite{ref:PLRevZ'} associated with the breaking of a $U(1)^{\prime}$ symmetry is predicted in many  
extensions of the Standard Model, such as the left-right symmetric model~\cite{ref:LRM}, Grand Unified Theories (GUT) based on SO(10)~\cite{ref:SO10} and E6~\cite{ref:E6, ref:E6Enrico}, and string inspired models~\cite {ref:strM}. If the $U(1)^{\prime}$ is flavor universal, it can only be the $U(1)_{\mbox{\tiny B-L}}$ in order to ensure the cancellation of all gauge anomalies. By allowing the SM fermions to have  generation dependent $U(1)^{\prime}$ charges, the $U(1)_{\nu}$, which differs from $U(1)_{\mbox{\tiny B-L}}$, may play the role of a  family symmetry which gives rise to fermion mass hierarchy and mixing patterns {\it \'a la} the Froggatt-Nielson mechanism~\cite{ref:frogNiel}. While~\cite{Ibanez:1994ig} found that to have realistic fermion mass and mixing patterns the $U(1)^{\prime}$ must be anomalous (and thus the $U(1)^{\prime}$ scale must be close to the string scale), more recent studies~\cite{ref:gauTrmNeuM,ref:SU(5)U(1), ref:highOpeWalter} have shown that realistic fermion mass and mixing patterns can arise from a non-anomalous $U(1)^{\prime}$ symmetry, and thus allowing the $U(1)^{\prime}$ scale to be as low as a TeV. In addition, all gauge anomalies can be cancelled with no exotic fields other than the right-handed neutrinos.

In the $U(1)_{\nu}$ model~\cite{ref:gauTrmNeuM}, which is based on the SM augmented with an additional generation-dependent, non- anomalous $U(1)_{\nu}$, small neutrino masses and their large mixing angles are generated at the TeV scale with all Yukawa coupling constants of the order of unity. Since the leptons are allowed to have generation dependent $U(1)_{\nu}$ charges in this model, the $Z^{\prime}$ gauge boson, which can be produced at the LHC, couples to the leptons non-universally. Even though the $Z^{\prime}$ couplings to the fermions are non-universal, all flavor changing neutral currents (FCNCs)~\cite{ref:FCNC} can still be satisfied. (More details are given in later sections.) In this paper, we study the LHC's discovery potential of the $Z^{\prime}$ in the $U(1)_{\nu}$ model. Specifically we concentrate on the capability of the LHC to distinguishing the $U(1)_{\nu}$ charges of the leptons, which is important in differentiating the generation-dependent $U(1)^{\prime}$ from the flavor universal case~\cite{ref:PLRevZ', ref:Z'LHCRizzo, ref:Z'LHCRizzoInd}. In addition to the searches of the resonance in the invariant mass distribution of the dilepton channels $Z^{\prime} \rightarrow \ell^{+}\ell^{-}$ ($\ell = e, \; \mu$), we also study the prospects of using the forward backward asymmetry distributions to distinguish different dilepton channels.

The paper is organized as follows. In Section~\ref{sec:model}, we briefly review the anomaly-free $U(1)_{\nu}$ model. Section~\ref 
{sec:searchZ'} shows the discovery potential of the $Z^{\prime}$ in the $U(1)_{\nu}$ model at the LHC as well as the capability of the LHC to distinguish different dilepton decay channels using the forward backward asymmetry distributions. In Section~\ref{subSec:bound}, a more general allowed parameter space for a generation dependent anomaly-free $U(1)$ model is given. Section~\ref{sec:conclusion} concludes the paper.

\section{\label{sec:model} The TeV Scale Seesaw Model With A Non-anomalous $U(1)_{\nu}$ Symmetry}

The $U(1)_\nu$ model extends the SM gauge group by an additional $U(1)$ symmetry. Because the SM fermions are allowed to have generation dependent $U(1)_{\nu}$ charges, the non-anomalous $U(1)_{\nu}$ symmetry can be different from $U(1)_{B-L}$. All gauge anomalies are cancelled in the model with no exotic fermions other than the three right-handed neutrinos, $\nu_{R_{t}}$ ($t = 1, 2, 3$). The $U(1)_{\nu}$ symmetry is broken at the TeV scale by the vacuum expectation value (VEV) of the scalar field, $\phi$, which has $U(1)_{\nu}$ charge of $+1$ and it is a singlet under the SM gauge group.  

In the conventional seesaw mechanism~\cite{ref:conSeeSaw}, the seesaw scale is generally on the order of $10^{14-16}$ GeV, which is close to the grand unification scale, making it inaccessible to the collider experiments. In the TeV scale seesaw model~\cite{ref:gauTrmNeuM} we consider, due to the $U(1)_{\nu}$ symmetry, the usual dimension-4 and dimension-5 operators for neutrino masses are forbidden; neutrino masses are generated by operators with higher dimensionality, thus allowing the seesaw scale to be lowered to a TeV. The complete Yukawa Lagrangian in the lepton sector is given as follows,
\begin{eqnarray}
\mathcal{L}_{\mbox{\tiny Yukawa}}  & = &   -\biggl[  \sum_{m,m^{\prime} =1}^3 \frac{\Gamma_{mm^{\prime}}^l}{\Lambda} \biggl( \frac{g_\phi}{\Lambda}\phi \biggr)^{q_{mm^{\prime}}} \overline{\ell^{c}}_{L_{m}} \ell_{L_{m^{\prime}}}HH 
\nonumber \\
& & + \sum_{m,t}^3 \Gamma_{mt}^{\nu} \biggl( \frac{g_\phi}{\Lambda}\phi \biggr)^{p_{mt}} \overline{\ell}_{L_{m}} \nu_{R_{t}} {H} 
\nonumber \\
& &
 + \sum_{t,t^{\prime}}^3 \Gamma_{tt^{\prime}}^n \biggl(\frac{g_\phi}{\Lambda}\phi\biggr)^{r_{tt^{\prime}}} \overline{\nu^{c}}_{R_{t}} \nu_{R_{t^{\prime}}} 
\nonumber\\
& & +
 \sum_{k,k^{\prime}}^3 \Gamma_{kk^{\prime}}^{e} \biggl( \frac{g_\phi}{\Lambda}\phi \biggr)^{s_{kk^{\prime}}} \overline{\ell}_{L_{k}} e_{R_{k^{\prime}}} \tilde{H} 
+ h.c. \biggr] \; ,
\label{eqn:LYukawa}
\end{eqnarray}
where $\ell_{L}$ and $e_{R}$ denote the left-handed and right-handed charged leptons, respectively, $H$ is the SM Higgs, and $\tilde{H} = i \sigma^2 H^{\dagger}$. The matrices $\Gamma_{mm^{\prime}}^l, \; \Gamma_{mt}^{\nu}, \;  \Gamma_{tt^{\prime}}^n, \; \Gamma_{kk^{\prime}}^{e}$ are the Yukawa couplings with $m, \; m^{\prime}, \;  t$ and $t^{\prime}$ being the generation indices for leptons, while $\Lambda$ is the cut-off scale of $U(1)_{\nu}$. When the $\phi$ field gets a VEV, the $U(1)_{\nu}$ symmetry is spontaneously broken, and all lepton masses are generated. Neutrino masses are suppressed by powers of $\epsilon \equiv g_{\phi} \frac{<\phi>}{\Lambda}$. If the powers are large enough, with $\epsilon \sim (0.1)$, the cutoff scale $\Lambda$ can be as low as a TeV.  

The resulting mass matrices in the lepton sector are characterized by the exponent matrices, which are determined by the fermions' $U(1)_\nu$ charges, $z_{f}$. Assuming that the quark and charged lepton masses are generated through dimension-4 operators, using gauge invariance and the six anomaly cancellation conditions, these mass matrices are determined by five independent parameters, $z_q, \; a, \; b, \; a^{\prime}$ and $b^{\prime}$~\cite{ref:gauTrmNeuM}. We further require that the second and third generations of charged leptons have the same charges and similarly for the right-handed neutrinos ($a' = 0, \; b' = 0$), as motivated by the maximal mixing between $\nu_{\mu}$ and $\nu_{\tau}$. Since the $\phi$ field is the only source of the $U(1)_\nu$ breaking, it follows that the Higgs charge $U(1)_{\nu}$ vanishes, $z_H = 0$. Consequently, there is no mixing between the SM $Z$ and the $Z^{\prime}$ at the tree level. The $Z-Z^{\prime}$ mixing can be induced at one-loop through the the self-energy diagram, $\Pi_{ZZ^{\prime}}$. This RG-induced $Z-Z^{\prime}$ gives rise to the following contribution~\cite{ref:rgeLoop} to the $\rho$ parameter
\begin{equation}
\Delta \rho \sim \frac{(\Delta M_{ZZ^{\prime}}^2)^2}{|M_{Z^{\prime}}^2-M_{Z}^2|M_{Z}^2} \sim \frac{1}{(16\pi^2)^2}\frac{(g_{z}Q_{z}g_{z^{\prime}}z_{f}m_{f}^2)^2}{|M_{Z^{\prime}}^2 - M_{Z}^2|M_{Z}^2} \; .
\end{equation}
Here the term $\Delta M_{ZZ^{\prime}}^{2}$ arises due to $\Pi_{ZZ^{\prime}}$;  it depends on 
the fermionic $Z$ couplings, $g_{z}Q_{z}$, the fermionic $Z^{\prime}$ coupling, $g_{z^{\prime}}z_{f}$, and the mass of the loop fermion, $m_{f}$. The electroweak precision measurements require extra contribution to $\rho$ parameter to be smaller than $\sim 10^{-4}$~\cite{ref:ewCons, ref:mixZZ, ref:reviewZ'}. While this in turn constrains our parameter space for $a$ and $b$, we have checked explicitly as shown in the subsequent sections that for $M_{Z^{\prime}} \sim 1$ TeV, all parameters considered in our model can satisfy the experimental precision electroweak constraints.
The $U(1)_{\nu}$ charges of various fields are summarized in Table~\ref{tbl:chrgZ}. 

The exponent matrices that parametrize the neutrino left-handed Majorana, Dirac, and right-handed Majorana mass matrices are given by,   
\begin{eqnarray}
\label{eqn:matx}
p & = & \left(\begin{array}{ccc} -2(a-b) & -(2a+b) & -(2a+b) \\ a+2b & a-b & a-b \\ a+2b & a-b & a-b \end{array} \right) \\
q & = & \left(\begin{array}{ccc} c+4a & c+a & c+a \\ c+a & c-2a & c-2a \\ c+a & c-2a & c-2a \end{array} \right) \\
r & = & \left(\begin{array}{ccc} c+4b & c+b & c+b \\ c+b & c-2b& c-2b \\ c+b & c-2b & c-2b \end{array} \right) \; ,
\end{eqnarray}
respectively, where $c = -\frac{2}{3}\frac{a^2+ab+b^2}{a+b}$. The exponent matrix for the charged lepton mass matrix is given by, 
\begin{equation}
s = \left(\begin{array}{ccc} 0 & -3a& -3a \\ 3a & 0 & 0 \\ 3a & 0 & 0 \end{array} \right) \; .
\end{equation}
Note that only elements which are integers are allowed in the exponent matrices, since the numbers of flavon field involved in the Froggatt-Nielson diagrams must be integers~\cite{ref:frogNiel}. Also for the case of negative integer elements, we insert $\phi^{\dagger}$ field which has the $U(1)_{\nu}$ charge of -1 instead of $\phi$ field defined in the Eq.~\ref{eqn:LYukawa}. By having non-integer exponents thus provides a way to generate texture zeros in the mass matrices. With this, neutrinos in this model can either be Dirac or Majorana fermions and both normal and inverted mass orderings can be accommodated. 

By allowing generation dependent lepton charges, tree-level FCNCs may arise through the (1, 2) and (1, 3) off-diagonal elements in the charged lepton mass matrix. In the case of $3a$ being an non-integer, all off-diagonal elements vanish and thus all tree-level FCNCs are avoided albeit having generation dependent lepton charges under the $U(1)_{\nu}$. Furthermore, an integral value for $|3a|$ is allowed as long as it is large enough to satisfy all experimental constraints on FCNCs, as the branching fractions of these processes, including $\mu^{+} \rightarrow e^{+}e^{-}e^{+}$, $\mu \rightarrow e$ conversion in nuclei, $\tau \rightarrow \ell \ell^{\prime} \ell^{\prime\prime}$, and $\tau \rightarrow \ell + $hadron(s), are 
roughly proportional to $(\epsilon^{|3a|})^{2}$. 
The gauge boson $Z^{\prime}$ couples to the fermions through
\begin{eqnarray}
\label{eqn:LZ'f}
\mathcal{L}_{\mbox{\tiny gauge}}  & = & -g_{z^{\prime}} \biggl[ \sum_{m=1}^3 (z_{q_{m}} \overline{q}_{L_{m}} \gamma^\mu Z_{\mu}^{\prime} q_{L_{m}} + z_{u_{m}} \overline{u}_{R_{m}} \gamma^\mu Z_{\mu}^{\prime} u_{R_{m}} 
\nonumber\\
& & 
+ z_{d_{m}} \overline{d}_{R_{m}} \gamma^\mu Z_{\mu}^{\prime} d_{R_{m}} + z_{\ell_{m}} \overline{\ell}_{L_{m}} \gamma^\mu Z_{\mu}^{\prime} {\ell}_{L_{m}} 
\nonumber\\
& & 
+ z_{e_{m}} \overline{e}_{R_{m}} \gamma^\mu Z_{\mu}^{\prime} e_{R_{m}}) 
+ \sum_{t=1}^3 z_{\nu_{t}} \overline{\nu}_{R_{t}} \gamma^\mu Z_{\mu}^{\prime} \nu_{R_{t}} \biggr]  \; ,
\end{eqnarray}
in which $g_{z^{\prime}}$ is the $U(1)_{\nu}$ gauge coupling constant. And the gauge couplings of the $Z^{\prime}$ to the fermions are determined by the $U(1)_{\nu}$ charges of the fermions. In the lepton sector, these charges also dictate the neutrino mixing patterns. Given that its mass $M_{Z^{\prime}} = \frac{g_{z^{\prime}}}{2} \left< \phi \right>$ is on the order of a TeV, the $Z^{\prime}$ can be produced at the LHC. The $U(1)_\nu$ model can thus be tested at the collider experiments through the dilepton decay channels.  

\begin{table}[tbh!]
\begin{tabular}{c|c}\hline\hline 
Field & $U(1)_{\nu}$ charge \\ [1ex] \hline 
$q_L, u_R, d_R$& $-\frac{1}{9} \frac{a^2+ab+b^2}{a+b}$ \\ [1ex]\hline 
$\ell_{L_{1}}, \; e_{R_{1}}$ & $\frac{-5a^2-5ab+b^2}{3(a+b)}$ \\ [1ex] \hline
$\ell_{L_{2}}, \ell_{L_{3}}, e_{R_{2}}, e_{R_{3}}$ & $\frac{(2a+b)^2}{3(a+b)}$ \\ [1ex]\hline
$\nu_{R_{1}}$& $\frac{a^2-5ab-5b^2}{3(a+b)}$ \\ [1ex] \hline 
$\nu_{R_{2}}, \nu_{R_{3}}$& $\frac{(a+2b)^2}{3(a+b)}$ \\ [1ex] \hline   
$H$& $0$ \\ [1ex] \hline
$\phi$& $1$  \\ [1ex] \hline\hline
\end{tabular}
\caption{The $U(1)_{\nu}$ charges of the SM fermions, SM Higgs field H, and the SM singlet flavon field, $\phi$.} 
\label{tbl:chrgZ} 
\end{table}

\section{Collider Signatures of the $U(1)_{\nu}$ Model}
\label{sec:searchZ'}


In the $U(1)_{\nu}$ model, the SM leptons can have generation dependent $U(1)_{\nu}$ charges, leading to non-universal couplings to the $Z^{\prime}$. We investigate the capability of various kinematic variables to distinguish the charged lepton $U(1)_{\nu}$ charges through the dilepton decay channels, $Z^{\prime} \rightarrow \ell^{+} \ell^{-}$. As a sample model point, we set $a = 13/3$ and $b = -5/3$, which gives quark and lepton $U(1)_{\nu}$ charges as summarized in Table~\ref{tbl:sChargeZ}.  In addition to being free of all gauge anomalies, these charges lead to realistic neutrino mass and mixing patterns, and satisfy all FCNC and electroweak precision measurement constraints. Furthermore, we take the gauge coupling parameter $g_{z^{\prime}} = 0.1$ in the analyses presented hereafter. In the example given above, neutrinos are Dirac particles since only the terms related to the matrix p contribute to the masses of neutrinos. The predicted absolute neutrino masses are, $m_3 \simeq 0$, $m_1 = 0.05$ eV, $m_2 = 0.05075$ eV with $\epsilon = 0.001$. These masses satisfy both experimental limits on $\Delta m_{atm}^{2}$ and $\Delta m_{\odot}^{2}$, and they are of the inverted hierarchical ordering. All tree-level FCNC processes are highly suppressed as these sample model parameters lead to a highly suppressed value for $\epsilon^{|6a|} = 0.001^{26} = 10^{-78}$. For $M_{Z^{\prime}} \sim 1$ TeV and $g_{Z^{\prime}} \sim 0.1$, the one-loop contribution to $Z-Z^{\prime}$ mixing for our bench mark model parameters is on the order of $\sim 10^{-4}$. This gives rise to $\Delta \rho$ approximately $\sim 10^{-9}$, which satisfies the current experimental constraints.

\begin{table}[tbh!]
\begin{tabular}{c|c}\hline\hline\
Field & $U(1)_\nu$ charge \\ \hline
$q_L, u_R, d_R$& $z_q = -\frac{43}{72}$ \\ \hline
$\ell_{L_{1}}, e_{R_{1}}$& $z_{\ell_{1}} = -\frac{55}{8}$ \\ \hline
$\ell_{L_{2,3}}, e_{R_{2,3}}$& $z_{\ell_{2}}  = z_{\ell_{3}} = \frac{49}{8}$ \\ \hline
$\nu_{R_{1}}$& $z_{\nu_{1}} = \frac{41}{8}$ \\ \hline
$\nu_{R_{2,3}}$& $z_{\nu_{2}} = z_{\nu_{3}} = \frac{1}{8}$ \\ \hline   
$H$& $z_H = 0$ \\ \hline
$\phi$& $z_{\phi} = 1$  \\ \hline \hline
\end{tabular}
\caption{The $U(1)_\nu$ charges of SM fermions, Higgs field H and scalar field $\phi$ in the specific case of $a = 13/3$, $b = -5/3$, and $g_{z^{\prime}} = 0.1$.}  
\label{tbl:sChargeZ}
\end{table}

\subsection{\label{subSec:invM} The $Z^{\prime}$ Discovery Potential at the LHC}


The $Z^{\prime}$ may be discovered by detecting excess signals from backgrounds near its resonance in the dilepton invariant mass distribution. We first calculate the leading order (LO) cross section of the exclusive $Z^{\prime}$ production, $PP(q\bar{q}) \rightarrow Z^{\prime} \rightarrow e^{+}e^{-}/\mu^{+}\mu^{-}$, at the LHC with the center of mass energy $\sqrt{s} = 14$ TeV. Since the $Z^{\prime}$ decay width is narrow, we neglect the interference term between the $Z^{\prime}$ and the SM gauge bosons. The cross section $\sigma(PP(q\bar{q}) \rightarrow Z^{\prime} \rightarrow e^{+}e^{-}/\mu^{+}\mu^{-})$ is given by,
\begin{widetext}
\begin{eqnarray}
\label{eqn:xSec1}
\sigma (PP(q\bar{q}) \rightarrow Z^{\prime} \rightarrow \ell^{+} \ell^{-}) 
& = & \frac{g_{z^{\prime}}^4z_q^2(z_{\ell_{m}})^2}{48\pi} 
\\
& & \times \sum_{q=u,d,c,s,b} \int_0^1 \!\! \int_0^1 \!\! \int_{-1}^1 \; dx_1 \; dx_2 \; d(\cos\hat{\theta}) 
f_q^{P}(x_1, Q^2) f_{\bar{q}}^{P}(x_2, Q^2)  \frac{\hat{s}(1+\cos^2\hat{\theta})}{(\hat{s}-M_{Z^{\prime}}^{2})^2+(M_{Z^{\prime}}\Gamma_{Z^{\prime}})^2} \; .
\nonumber
\end{eqnarray}
Taking $Q^2 = M_{Z^{\prime}}^2$ and integrating out $\cos\hat{\theta}$, with $\hat{\theta}$ being the angle between the incoming annihilating quark and outgoing negatively charged lepton in the dilepton center of mass frame, the total cross section is simplified to
\begin{equation}
\label{eqn:xSec2}
\sigma(PP(q\bar{q}) \rightarrow Z^{\prime} \rightarrow \ell^{+} \ell^{-}) =  \frac{g_{z^{\prime}}^4z_q^2(z_{\ell_{m}})^2}{18\pi} 
\! \! \!
\sum_{q=u,d,c,s,b} \int_0^1 \!\! \int_0^1 \,dx_1 \,dx_2 
f_q^{P}(x_1, M_{Z^{\prime}}^2)f_{\bar{q}}^P(x_2,M_{Z^{\prime}}^2) \frac{\hat{s}}{(\hat{s}-M_{Z^{\prime}}^2)^2+(M_{Z^{\prime}} \Gamma_{Z^{\prime}})^2} \; , 
\end{equation}
and the total $Z^{\prime}$ decay width becomes 
\begin{equation}
\label{eqn:dWidth}
\Gamma_{Z^{\prime}} =  \frac{g_{z^{\prime}}^{2} M_{Z^{\prime}}}{36\pi} \biggl[ 23 \biggl( \frac{a^2+ab+b^2}{3(a+b)} \biggr)^2 
+ \beta_t \biggl(1+2 \frac{m_t^2}{M_{Z^{\prime}}^2} \biggr) \biggl(\frac{a^2+ab+b^2}{3(a+b)} \biggr)^2 + 27 a^2 + 9 b^2 \biggr] \; ,
\end{equation}
\end{widetext}
where $\beta_t = \sqrt{1 - 4 m_t^2 /  M_{Z^{\prime}}^2}$ and $m_t$ is the top quark mass.
In Eq.~\ref{eqn:xSec1} and \ref{eqn:xSec2}, $f_q^{P}(x_1, Q^2)$ and $f_{\bar{q}}^{P}(x_2, Q^2)$ are parton distribution functions for the protons with $\hat{s} = x_1 x_2 s$. For our choice of parameters, neutrinos are Dirac particles and thus the right-handed neutrinos in this model can be very light. This allows the $Z^{\prime}$ to decay into right-handed neutrino pairs; this decay channel has been included, along with the decay channel of $Z^{\prime}$ into the SM fermions,  in the total decay width given by Eq.~\ref{eqn:dWidth}.


The current experimental limit for $M_{Z^{\prime}}$ is $\sim$ (900 GeV-1 TeV)~\cite{ref:CDFz'Con1, ref:CDFz'Con2, ref:D0z'}. While CDF Run II has discovered at $2.5\sigma$ significance an excess at the dielectron invariant mass of 240 GeV~\cite{ref:CDF240}, this excess has not been  confirmed by D$\O$ nor has a similar excess been discovered in the dimuon channel. To be conservative, we consider in our analyses $M_{Z^{\prime}} \geq 1$ TeV.

The main backgrounds which can mimic the signal events $Z^{\prime} \rightarrow e^{+}e^{-}$ and $Z^{\prime} \rightarrow \mu^{+}\mu^{-}$, can be categorized into two types, the reducible and irreducible backgrounds. Without any selection cuts, the dominant QCD backgrounds are inclusive jets, $W + $jets, $W+\gamma$, $Z + $jets, $Z+\gamma$, $\gamma + $jets, and $\gamma+\gamma$, which are caused by jets or photons faking the electrons or muons. These backgrounds can be reduced by applying lepton identification criteria and an isolation cut, which reject fake electrons with a rejection factor $R_{e-jet} = 4 \times 10^3$ in the case of jets and $R_{e-\gamma} = 10$ in the case of photons~\cite{ref:ATLASNote}. The QCD jet backgrounds can be further reduced by applying additional geometrical acceptance of $|\eta| < 2.5$~\cite{ref:ATLASTDR} and by imposing the transverse momentum cuts~\cite{ref:ATLASNote} which require $P_{t} \geq 65$ GeV for electrons and $P_{t}  \geq 30$ GeV for muons. Since the two leptons are more likely to fly back to back in the detector, we also require the azimuthal angle between the two outgoing leptons to be $|\Delta \phi_{\ell \ell }| > \frac{\pi}{2}$ (though this does not affect either the signals or backgrounds most of the time). After these selection cuts, which are summarized in Table~\ref{tbl:selCrit}, the dominant backgrounds are the irreducible backgrounds, which mainly come from the SM Drell-Yan processes. Other processes like decay products from $WW$, $WZ$, $ZZ$ and $t \bar{t}$ etc., can be ignored since their cross sections are very small. Hence, in our study, $Z^{\prime} \rightarrow e^{+}e^{-}$ and $Z^{\prime} \rightarrow \mu^{+} \mu^{-}$ and only the SM Drell-Yan processes are simulated in PYTHIA~\cite{ref:pythia}. To generate the SM Drell-Yan processes more efficiently, we also require $\hat{m} = \sqrt{\hat{s}} > 500$ GeV. 

\begin{figure}[tbh!]
\includegraphics[scale=0.8, angle = 90, width = 80mm, height = 50mm]{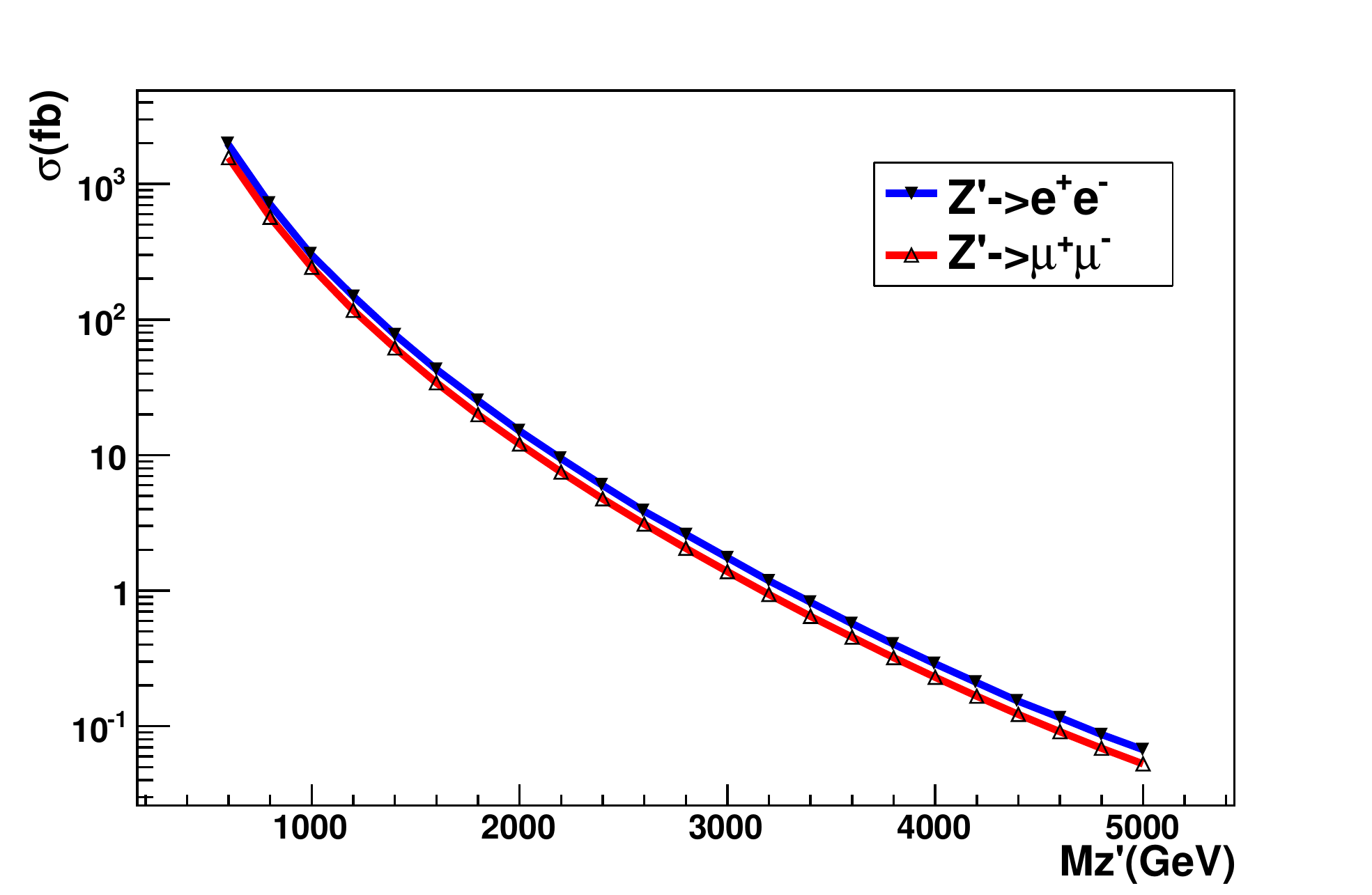}
	\caption{Cross Sections of $Z^{\prime} \rightarrow e^{+}e^{-}$ and $Z^{\prime} \rightarrow \mu^{+}\mu^{-}$ as a function of the $Z^{\prime}$ mass.}
	\label{fig:xSectEM}
\end{figure}

\begin{figure}[tbh!]
\includegraphics[scale=0.8, angle = 90, width = 80mm, height = 50mm]{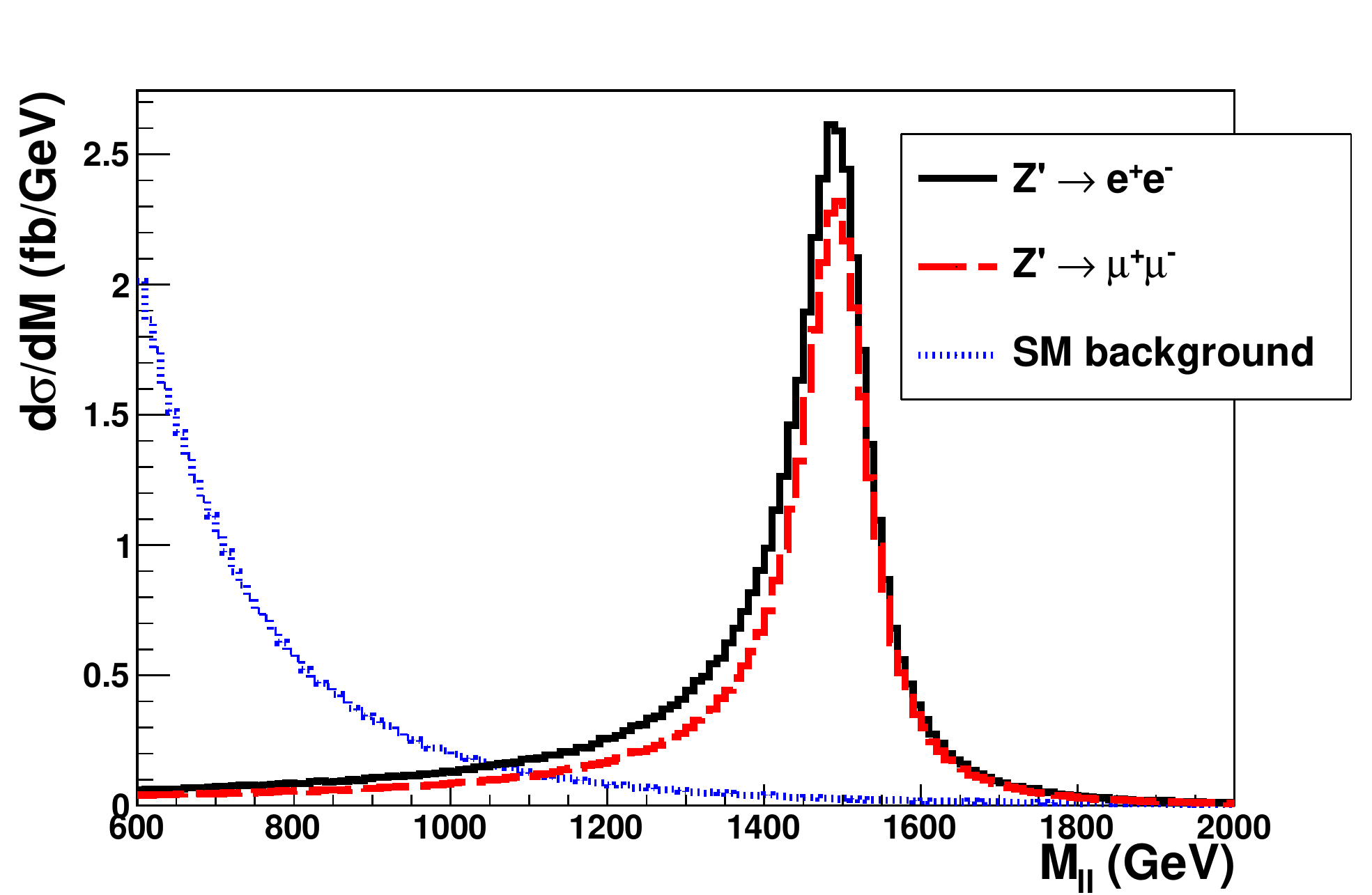}
    \caption{Dilepton invariant mass distributions $M_{Z^{\prime}} = 1.5$ TeV at $\sqrt{s} = 14$ TeV. The black solid line represents the $Z^{\prime} \rightarrow e^{+}e^{-}$ channel, while the red dashed line is the $Z^{\prime} \rightarrow \mu^{+}\mu^{-}$ channel. The blue dotted line is the SM backgrounds for both channels, which are almost identical for both channels at the tree level since $Z$ and $\gamma$ couple to SM fermions universally. }
    \label{fig:invM}
\end{figure}

With these selection cuts, Fig.~\ref{fig:xSectEM} shows the the cross sections of $Z^{\prime} \rightarrow e^{+}e^{-}$ and $Z^{\prime} \rightarrow \mu^{+}\mu^{-}$ in the $U(1)_{\nu}$ model as a function of the $Z^{\prime}$ mass. The dilepton invariant mass distributions in the $e^{+}e^{-}$ and $\mu^{+}\mu^{-}$ channels for $M_{Z^{\prime}} = 1.5$ TeV are shown in Fig~\ref{fig:invM}. The SM backgrounds, which are almost identical for both channels at the tree level since the gauge couplings of $Z$ and $\gamma$ to the SM fermions are universal, are highly suppressed in the $Z^{\prime}$ resonance region, allowing a clear distinction between the signals from the backgrounds.

\begin{table}[tbh!]
\begin{tabular}{c|c} \hline
$Z^{\prime} \rightarrow e^{+}e^{-}$& $Z^{\prime} \rightarrow \mu^{+}\mu^{-}$ \\ \hline 
$P_{t} (e) > 65$ GeV & $P_{t} (\mu) > 30$ GeV \\ \hline
\multicolumn{2}{c}{$|\Delta \phi_{\ell \ell}| > \frac{\pi}{2}$} \\ \hline
\multicolumn{2}{c}{$\eta(\ell) < 2.5$} \\ \hline
\end{tabular}
\caption{Selection criteria for the $Z^{\prime} \rightarrow e^{+}e^{-}$ and $Z^{\prime} \rightarrow \mu^{+}\mu^{-}$ decay channels.}
\label{tbl:selCrit}
\end{table}

\begin{figure}[tbh!]
\includegraphics[scale=0.8, angle = 90, width = 80mm, height = 50mm]{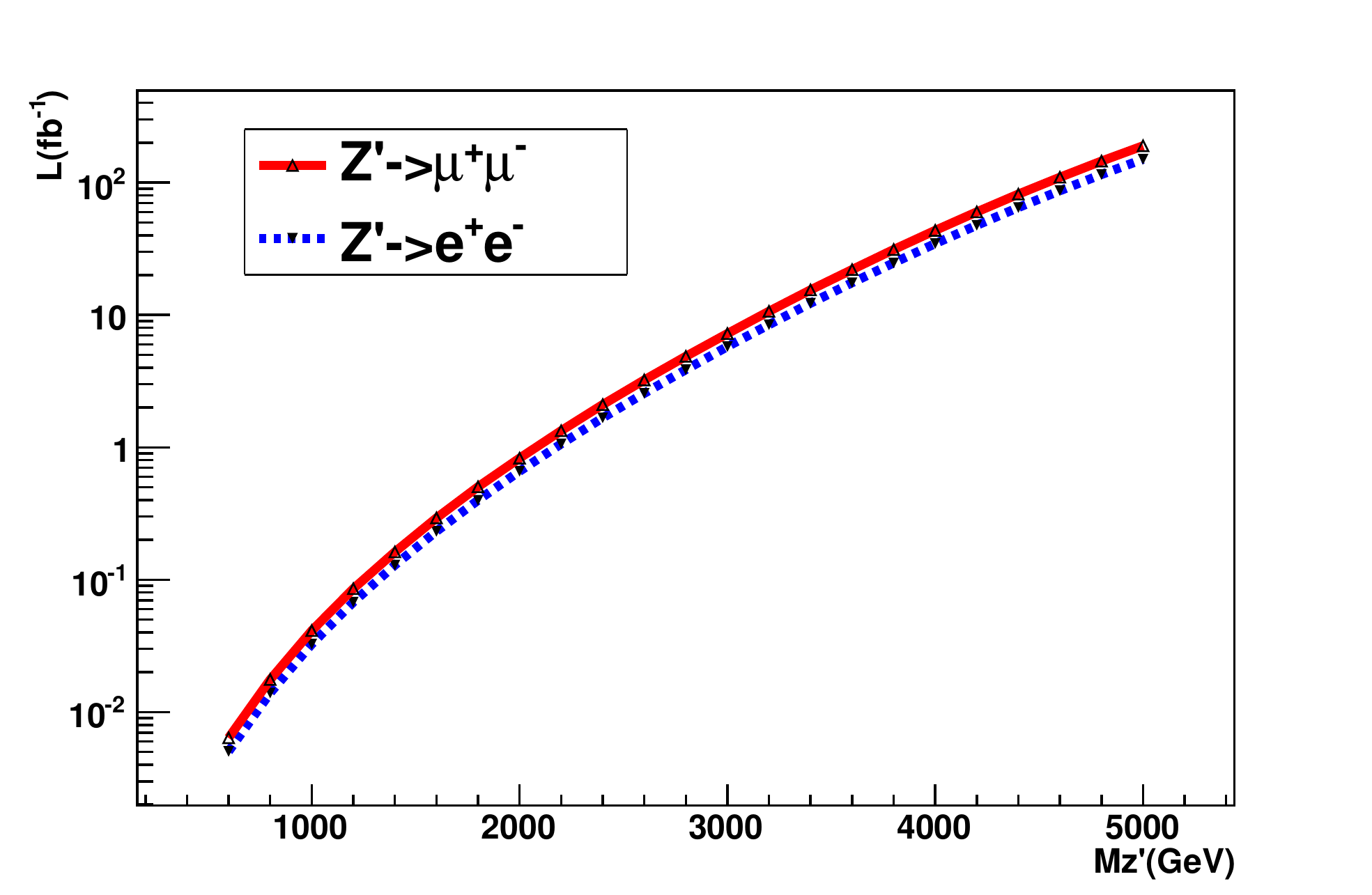}
	\caption{Integrated luminosity required for a $5\sigma$ $Z^{\prime}$ discovery, with $S/\sqrt{B} > 5$ and $S > 10$ at $\sqrt{s} = 14$ TeV.}
	\label{fig:lumi}
\end{figure}

To deduce the $Z^{\prime}$ discovery potential, we use the usual counting method. Let $S$ be the number of $Z^{\prime}$ signal events, which satisfies $|M_{\ell \ell} - M_{Z^{\prime}}| < 2\Gamma_{Z^{\prime}}$ and $B$ be the number of SM Drell-Yan background events. Both $S$ and $B$ satisfy the aforementioned selection cuts. One can then define the variable  $S/\sqrt{B} \equiv \sigma_{S}*L/\sqrt{\sigma_{B}*L}$, where $\sigma_{S}$ and $\sigma_{B}$ respectively are the total cross sections of the signal and background events, and $L$ is the integrated luminosity. A $5\sigma$ $Z^{\prime}$ discovery corresponds to $S/\sqrt{B} > 5$ while a $2\sigma$ exclusion corresponds to $S/\sqrt{B} < 2$. With $100$ fb$^{-1}$ of integrated luminosity and center of mass energy $\sqrt{s} = 14$ TeV,  there are 4151 signal events and 102 Drell-Yan background events for the case of $M_{Z^{\prime}} = 1.5$ TeV in the dielectron channel. For the dimuon channel, the numbers of signal events and background events are 3104 and 108, respectively. In the case of $\sqrt{s}=10$ TeV and $M_{Z^{\prime}} = 1$ TeV, with $200$ pb$^{-1}$ of data,  there are 20 signal events in the dielectron channel and $\sim 17$ signal events in the dimuon channel, both with less than 1 event from the SM backgrounds. Fig.~\ref{fig:lumi} shows the integrated luminosity required for a $5\sigma$ $Z^{\prime}$ discovery through both the electron and muon channels at $\sqrt{s} = 14$ TeV. Since the counting method works for a large amount of events which satisfy the Gaussian distribution, we further require the number of signal events $S > 10$. Therefore, with $100$ fb$^{-1}$ of data, the LHC can discover a $Z^{\prime}$ with a mass up to 4.5 TeV at $\sqrt{s} = 14$ TeV.

\subsection{\label{subSec:BR} Establishing the $Z^{\prime}$ Flavor Dependence with Decay Branching Fractions}

\begin{figure}[tbh!]
\includegraphics[scale = 0.8, angle = 90, width = 80mm, height = 50mm]{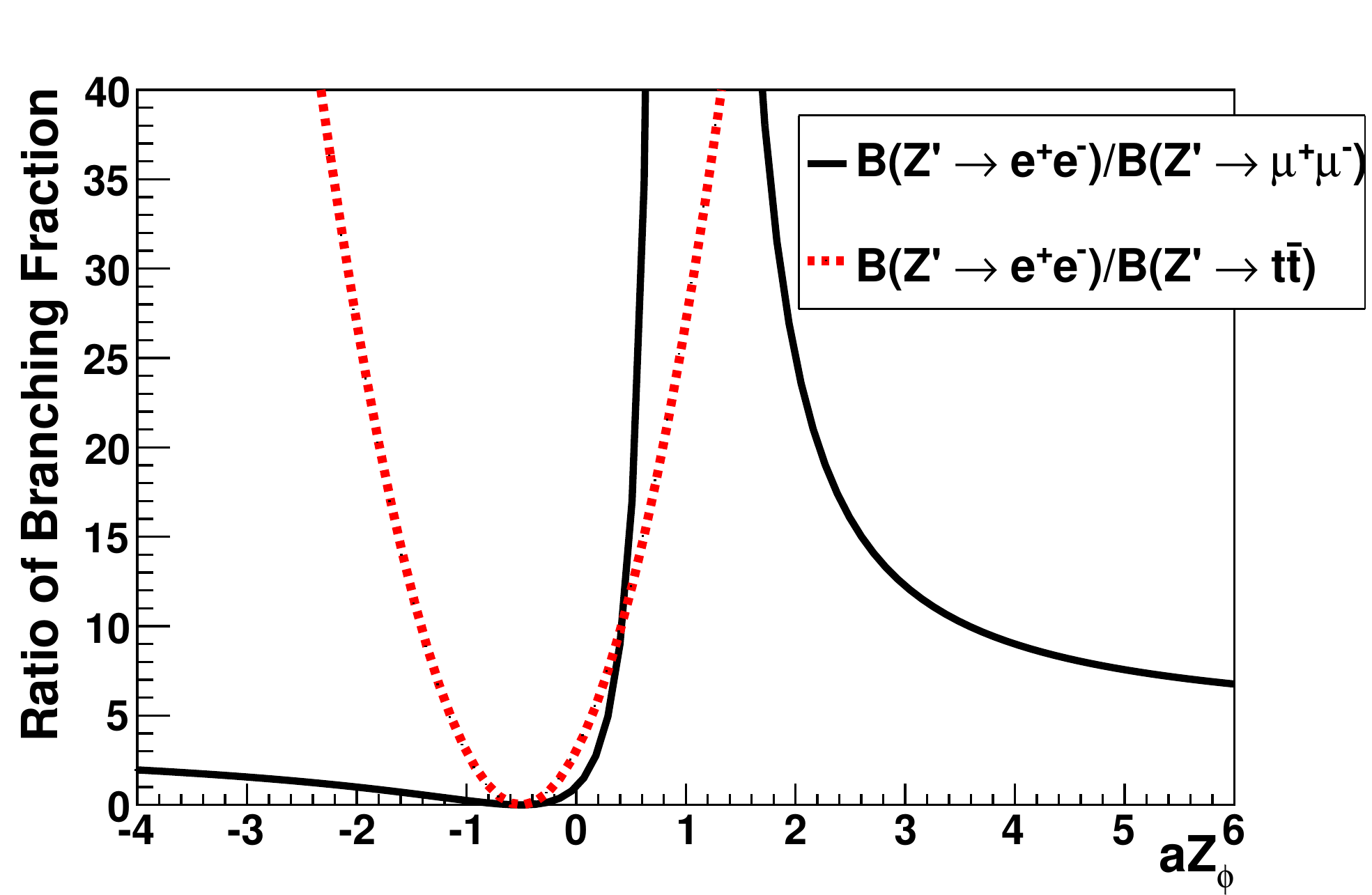}
	\caption{Ratios of branching fractions ($R_{e\mu}$ and $R_{et}$) as a function of variable $Z_{\phi} = \frac{-3(a+b)}{a^{2}+ab+b^{2}}$.}
	\label{fig:brr}
\end{figure}

While there is a great potential for the $Z^{\prime}$ discovery at the LHC during its early low luminosity running as shown in the previous section, to establish the flavorful nature of the $Z^{\prime}$, after it has been discovered in the dilepton channels, turns out to be more difficult. Since in the $U(1)_{\nu}$ model, the electron and muon are allowed to have different $U(1)_{\nu}$ charges, the ratio of the decay branching fractions, $R_{e\mu} = \frac{B(Z^{\prime} \rightarrow e^{+}e^{-})}{B(Z^{\prime} \rightarrow \mu^{+}\mu^{-})}$, can in general differ from $1$. The ratio $R_{e\mu}$ depends on the parameter, $Z_{\phi} = \frac{-3(a+b)}{a^{2}+ab+b^{2}}$, which is given in terms of the charge parameters $a$ and $b$, as  
\begin{equation}
R_{e\mu} = \frac{B(Z^{\prime} \rightarrow e^{+}e^{-})}{B(Z^{\prime} \rightarrow \mu^{+} \mu^{-})} 
= \biggl(\frac{z_{\ell_1}}{z_{\ell_2}}\biggr)^2 
= \biggl( \frac{1+2aZ_{\phi}}{1-aZ_{\phi}} \biggr)^2 \; .
\end{equation}
If the $Z^{\prime}$ is also discovered in the $t\overline{t}$ channel, by measuring $R_{e\mu}$ and the ratio of the decay branching fractions of $e^{+}e^{-}$ to $t\overline{t}$ channels, 
\begin{equation}
R_{et} = \frac{B(Z^{\prime} \rightarrow e^{+}e^{-})}{B(Z^{\prime}  \rightarrow t \bar{t})} = (\frac{z_{\ell_1}}{z_q})^2 = 3(1+2aZ_{\phi})^2 \; , 
\end{equation}
a unique determination of $aZ_{\phi}$ can be obtained (see Fig.~\ref{fig:brr}).

It is to be noted that when $aZ_{\phi} = -1/2$, the branching fraction of $Z^{\prime} \rightarrow e^{+}e^{-}$ vanishes, as the electron is not charged under the $U(1)_{\nu}$ in this case. Consequently, if the $Z^{\prime}$ is discovered in the dimuon and (or) $t\bar{t}$ channels, and if there is no evidence in the dielectron channel, this then suggests that the $U(1)^{\prime}$ symmetry is flavor dependent. Similarly, if $aZ_{\phi}=1$, then the left-handed muon is not charged under the $U(1)^{\prime}$. Thus one would expect a discovery in the dielectron channel, but not in the dimuon channel.  In principle, if we can detect and measure all  $Z^{\prime}$ decay channels precisely enough, we can determine the parameters $a$ and $b$, which are the parameters that predict neutrino mixing angles, mass ordering and their Dirac versus Majorana nature. 

\begin{figure}[tbh!]
\includegraphics[scale=0.8, angle = 90, width = 80mm, height = 50mm]{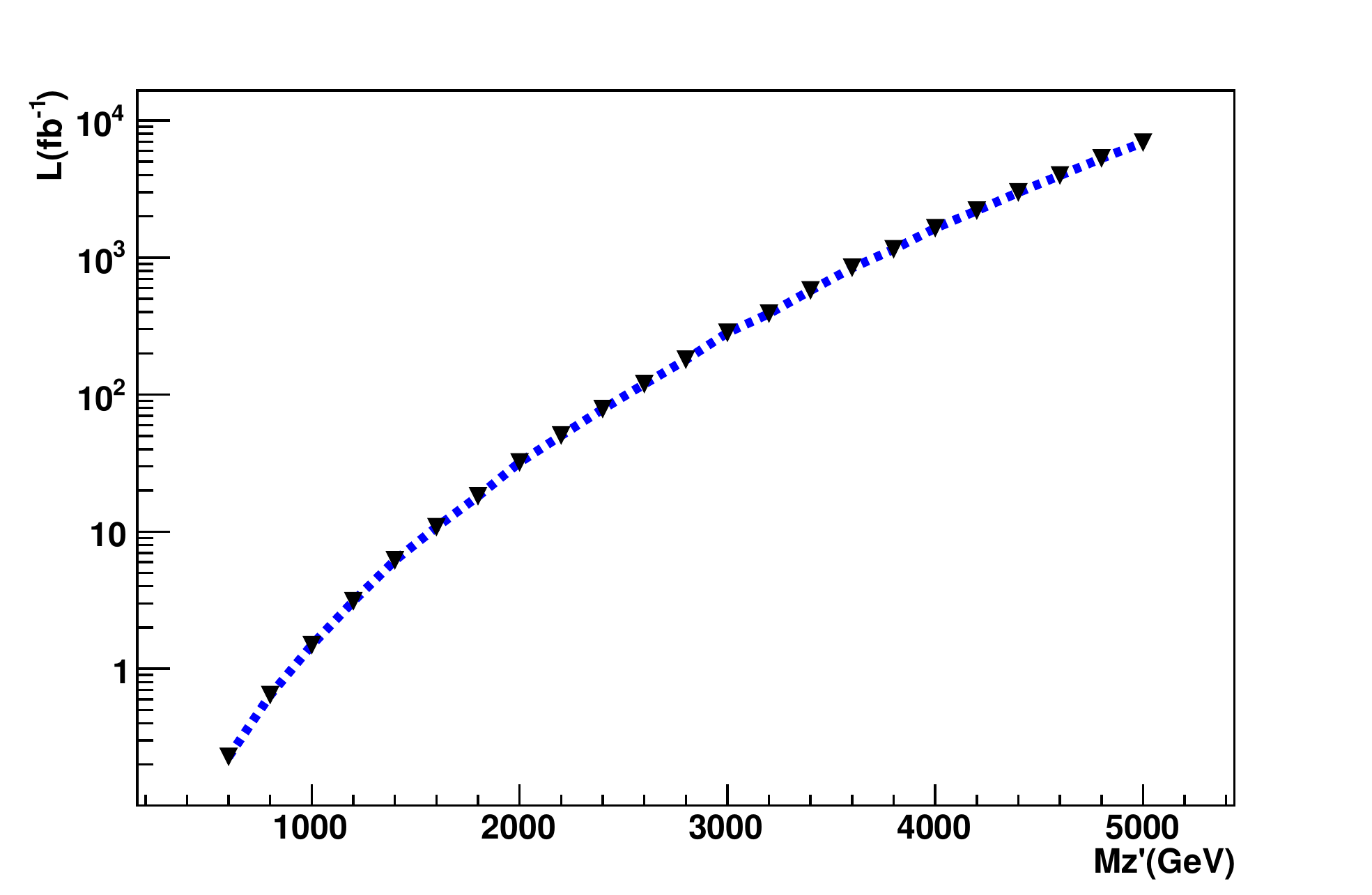}
    \caption{Integrated luminosities required at $\sqrt{s} = 14$ TeV to distinguish the dielectron and dimuon channels at $5 \sigma$ significance for different $Z^{\prime}$ mass assumptions.}
    \label{fig:lumiR}
\end{figure}

Our bench mark point, $z_{\ell_{1}} = -55/8$ and $z_{\ell_{2}} = 49/8$, predicts a ratio of the branching fractions of $R_{e\mu} = (55/49)^2 \approx 1.26$. To study the integrated luminosity required for distinguishing the $e^{+}e^{-}$ and $\mu^{+}\mu^{-}$ channels at $5\sigma$ using the counting method, we define the following variable, 
\begin{equation}
\frac{D}{\sqrt{B}} = \frac{\sigma_{ee}\ast L - \sigma_{\mu\mu} \ast L}{\sqrt{\sigma_{\mu\mu} \ast L}} \; ,
\end{equation}
where $\sigma_{ee}$ and $\sigma_{\mu\mu}$ are the cross sections of the $e$ and $\mu$ events. The parameter $D$ thus gives the difference in the numbers of events between $Z^{\prime} \rightarrow e^{+}e^{-}$ and $Z^{\prime} \rightarrow \mu^{+}\mu^{-}$.  If $D/\sqrt{B} > 5$, the difference between the dielectron and dimuon events is above $5 \sigma$ significance. Similar to the requirement of $S>10$ discussed previously, we also require $D > 10$. The integrated luminosity required as a function of the $Z^{\prime}$ mass is shown in Fig.~\ref{fig:lumiR}. At $\sqrt{s} = 14$ TeV  with $500$ fb$^{-1}$ of data, a statistically significant distinction between the branching fractions for $e$ and $\mu$ channels can be obtained up to $M_{Z^{\prime}} = 3$ TeV. For the case of $\sqrt{s} = 10$ TeV, to distinguish the dielectron and dimuon channels, more data are needed. For $M_{Z^{\prime}} = 1.0$ TeV, to get $D/\sqrt{B} > 5$ and $D > 10$, an integrated luminosity of $8.32$ fb$^{-1}$ is needed.

\subsection{\label{subSec:AFB} Establishing the $Z^{\prime}$ Flavor Dependence with Forward Backward Asymmetry Distributions}


Even though the flavorful nature of the $Z^{\prime}$ can be determined by measuring the ratio of branching fractions, $R_{e\mu}$, as discussed above, since the ratio of branching fractions depends on the square of the charged lepton $U(1)_{\nu}$ charges, the measurement of $R_{e\mu}$ cannot tell the relative sign between $z_{\ell_{1}}$ and $z_{\ell_{2}}$. On the other hand, the sign of the $U(1)_{\nu}$ charge can be determined by measuring the forward backward asymmetry distribution, $A_{\mbox{\tiny FB}}$, as $A_{\mbox{\tiny FB}}$ is due to the interference between the $Z^{\prime}$ and the SM gauge bosons, Z and $\gamma$, which can be small, especially near the $Z^{\prime}$ resonance as discussed in Section~\ref{subSec:invM}. Hence, this measurement requires a large amount of data, which may be possible at the SLHC~\cite{ref:LHCRunPlan} allowing an instantaneous luminosity of $10^{35 }$ cm$^{-2}$s$^{-1}$. 

The forward backward asymmetry, $A_{\mbox{\tiny FB}}$, is defined as,
\begin{equation}
A_{FB} \equiv \frac{\sigma_{F} - \sigma_{B}}{\sigma_{F} + \sigma_{B}} \; ,
\end{equation}
where the total cross sections of the forward and backward events are given by
\begin{eqnarray}
\sigma_{F} & = & \int_0^1 \, d\cos\theta \frac{d \sigma(P\bar{P} \rightarrow l^{+}l^{-})}{d \cos\theta} \; , \\
\sigma_{B} & = & \int_{-1}^0 \, d\cos\theta \frac{d \sigma(P\bar{P} \rightarrow l^{+}l^{-})}{d \cos\theta} \; ,
\end{eqnarray}
and the angle $\theta$ is defined as the angle between the direction of the outgoing negatively charged lepton with respect to the direction of the incoming annihilating quark in the lab frame. Note that this is different from the usual $\hat{\theta}$ angle which is the angle defined in the center of mass frame, and the angles $\theta$ and $\hat{\theta}$ are related by
\begin{equation}
\frac{d \sigma(P\bar{P} \rightarrow l^{+}l^{-})}{d \cos\theta} 
= \frac{d \sigma(P\bar{P} \rightarrow l^{+}l^{-})}{d \cos\hat{\theta}} \frac{d \cos\hat{\theta}}{d \cos\theta} \; .
\end{equation}
We can thus calculate the cross section in the center of mass frame (the general formula can be found in~\cite{ref:AFBxSec}) and transform it from the center of mass frame to the lab frame (the formula can be found in~\cite{ref:transf}). 

Due to an intrinsic ambiguity in deciding which proton the quark comes from, it is still tricky to determine $\cos\theta$. Inspired by~\cite{ref:cosThExp, ref:CMSAFBNote}, we therefore define $\cos\theta^{\ast}$ with $\theta^{\ast}$ being the angle between the direction of the negatively charged lepton with respect to the direction of the proton which contributes the annihilating quark in the lab frame. Since experimentally, it is hard to determine which proton contributes to the quark, due to the fact that a  quark in the proton typically carries a larger momentum fraction than does an antiquark, we assume that the longitudinal motion of the dilepton system is in the direction of the proton which contributes the annililating quark in the lab frame~\cite{ref:CMSAFBNote}.

\begin{figure}[tbh!]
\includegraphics[scale = 0.8, angle = 90, width = 80mm, height = 50mm]{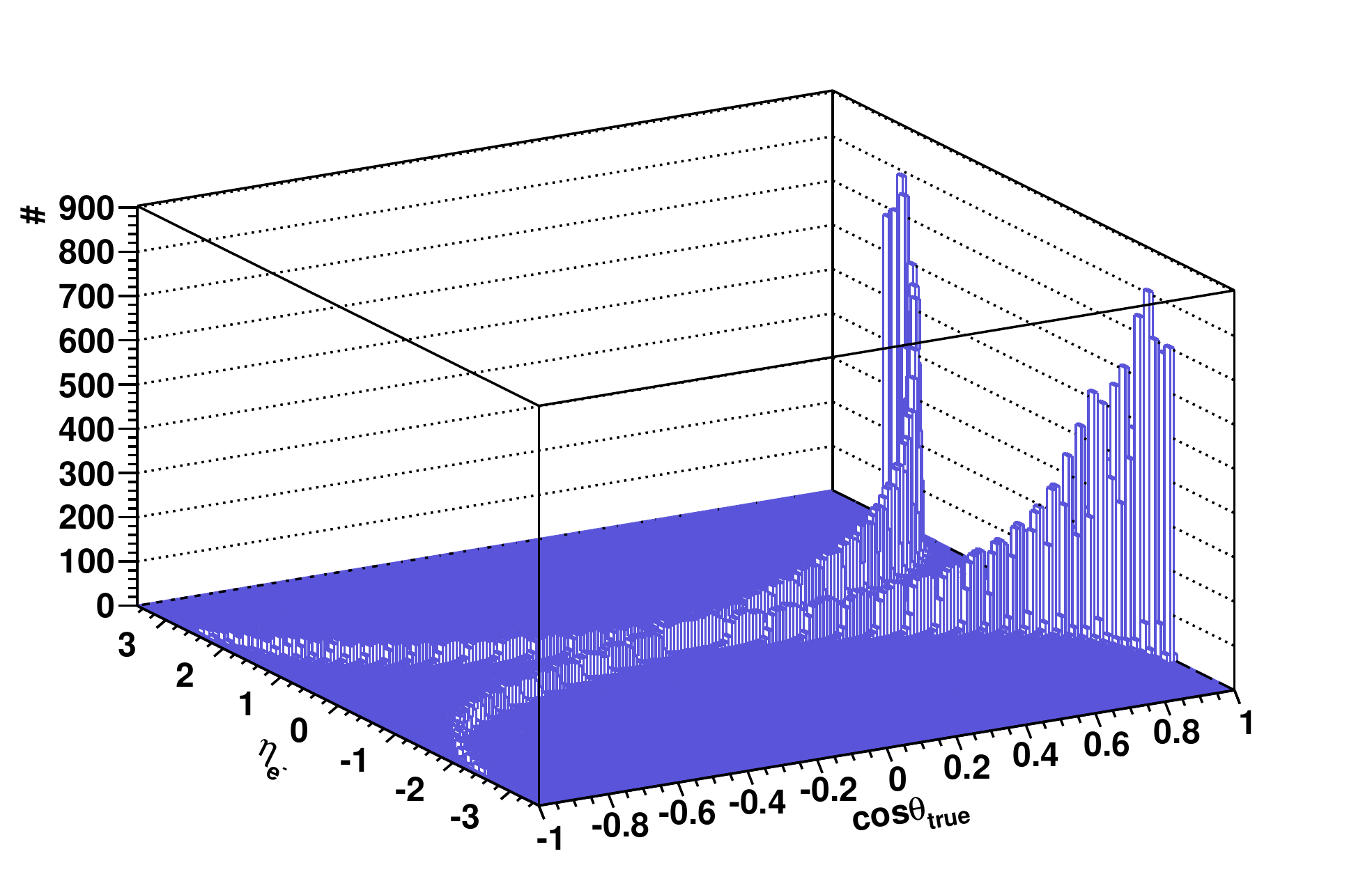}
	\caption{The lego plot of $\cos \theta_{\mbox{\tiny true}}$ versus $\eta$ in the dielectron channel. (The dimuon channel exhibits very similar correlation.)}
	\label{fig:cosThetaEtaTruth}
\end{figure}

\begin{figure}[tbh!]
\includegraphics[scale = 0.8, angle = 90, width = 80mm, height = 50mm]{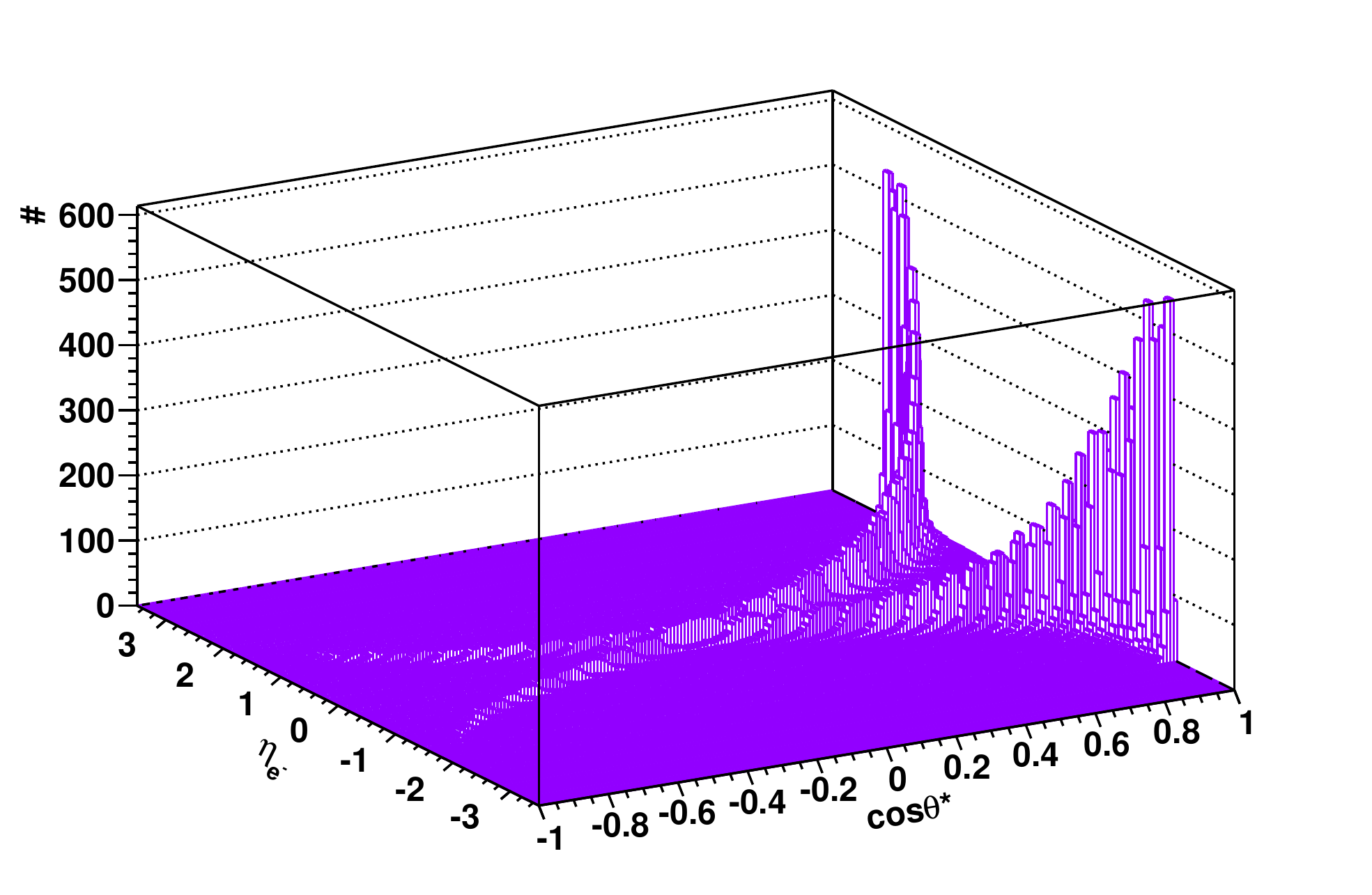}
	\caption{The lego plot of $\cos \theta^{\ast}$ versus  $\eta$ in the dielectron channel. (The dimuon channel exhibits very similar correlation.)}
	\label{fig:cosThetaEtaExp}
\end{figure}

\begin{figure}[tbh!]
\includegraphics[scale = 0.8, angle = 90, width = 80mm, height = 50mm]{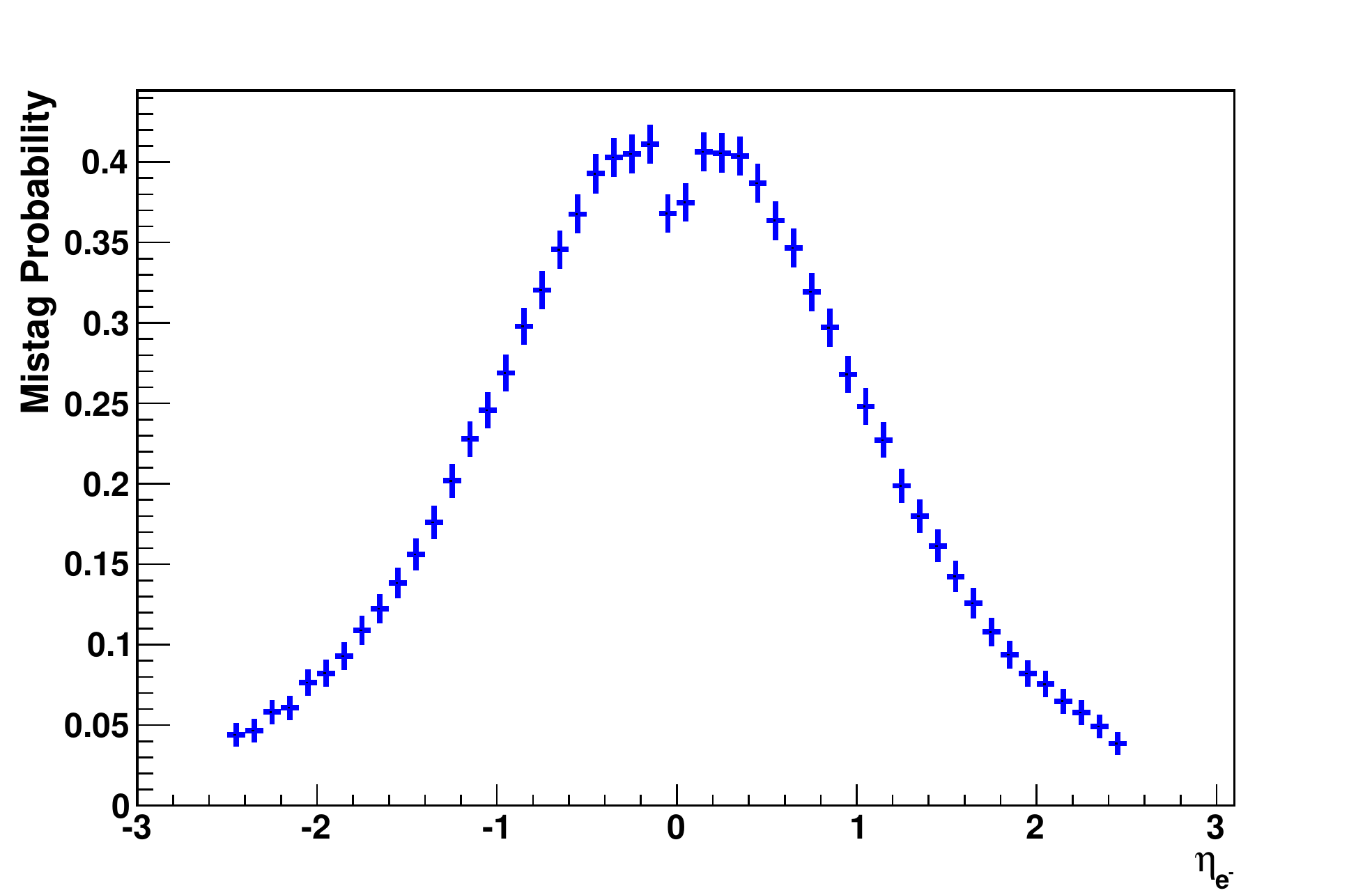}
	\caption{Mistagging rate as a function of $\eta_{e^{-}}$.}
	\label{fig:misTagEta}
\end{figure}

To understand how well $\cos \theta^{*}$ describes the variable $\cos\theta$, we compare the lego plot of $\cos \theta_{\mbox{\tiny true}}$ versus $\eta$ (Fig.~\ref{fig:cosThetaEtaTruth}) and the lego plot of $\cos \theta^{\ast}$ versus $\eta$ (Fig.~\ref{fig:cosThetaEtaExp}), where $\theta_{\mbox{\tiny true}}$ is the Monte Calo true angle between the direction of the negatively charged lepton and that of the annihilating quark. 
In these two figures, the events generated include the full interference structure of $\gamma$/Z/$Z^{\prime}$ and satisfy the transverse momentum cut $P_{t}(e) > 65$ GeV or $P_{t}(\mu) > 30$ GeV, back to back selection cut $|\Delta \phi_{\ell\ell}| > \frac{\pi}{2}$, and the requirement that the invariant mass of the dilepton system is within the range of 500 GeV to 2000 GeV. We find that $\cos \theta^{\ast}$ can be used as a good approximation for $\cos \theta_{\mbox{\tiny true}}$. Although those two figures look very similar, the mistagging rate (that is, $\cos \theta^{*}$ having a wrong sign compared to $\cos \theta_{\mbox{\tiny true}}$) is still high at the low $\eta$ region, as shown in Fig.~\ref{fig:misTagEta}. With an additional cut of $|\eta| > 0.8$, the mistagging rate can be reduced to a reasonable level ($\sim 30 \%$) without losing too much data. 
Therefore, we use $\cos \theta^{\ast}$ to approximate $\cos \theta_{\mbox{\tiny true}}$ when extracting the forward backward asymmetry in the simulations with the requirement of $0.8 < |\eta| < 2.5$.

%
%

\begin{figure}[tbh!]
\includegraphics[scale = 0.8, angle = 90, width = 80mm, height = 50mm]{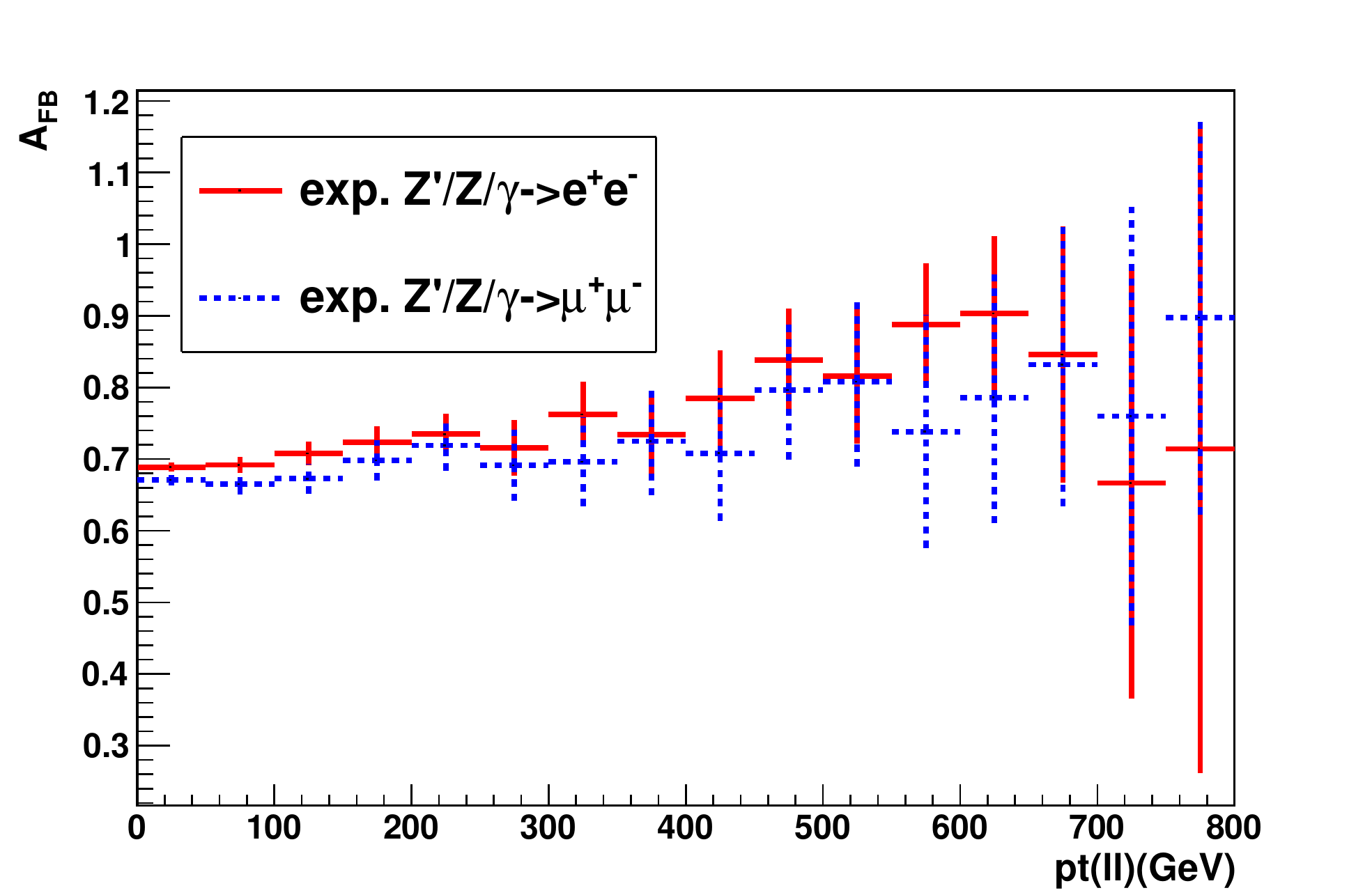}
	\caption{The forward backward asymmetry distribution as a function of the dilepton transverse momentum for $M_{Z^{\prime}} = 1.5$ TeV in the case of $\sqrt{s} = 14$ TeV with $500$ fb$^{-1}$ of data. The red lines represent the electron channel and the dotted blue lines denote the muon channel. The error bars are the statistical uncertainties normalized to $500$ fb$^{-1}$ of data.}
	\label{fig:AFBvsPt}
\end{figure}

\begin{figure}[tbh!]
\includegraphics[scale = 0.8, angle = 90, width = 80mm, height = 50mm]{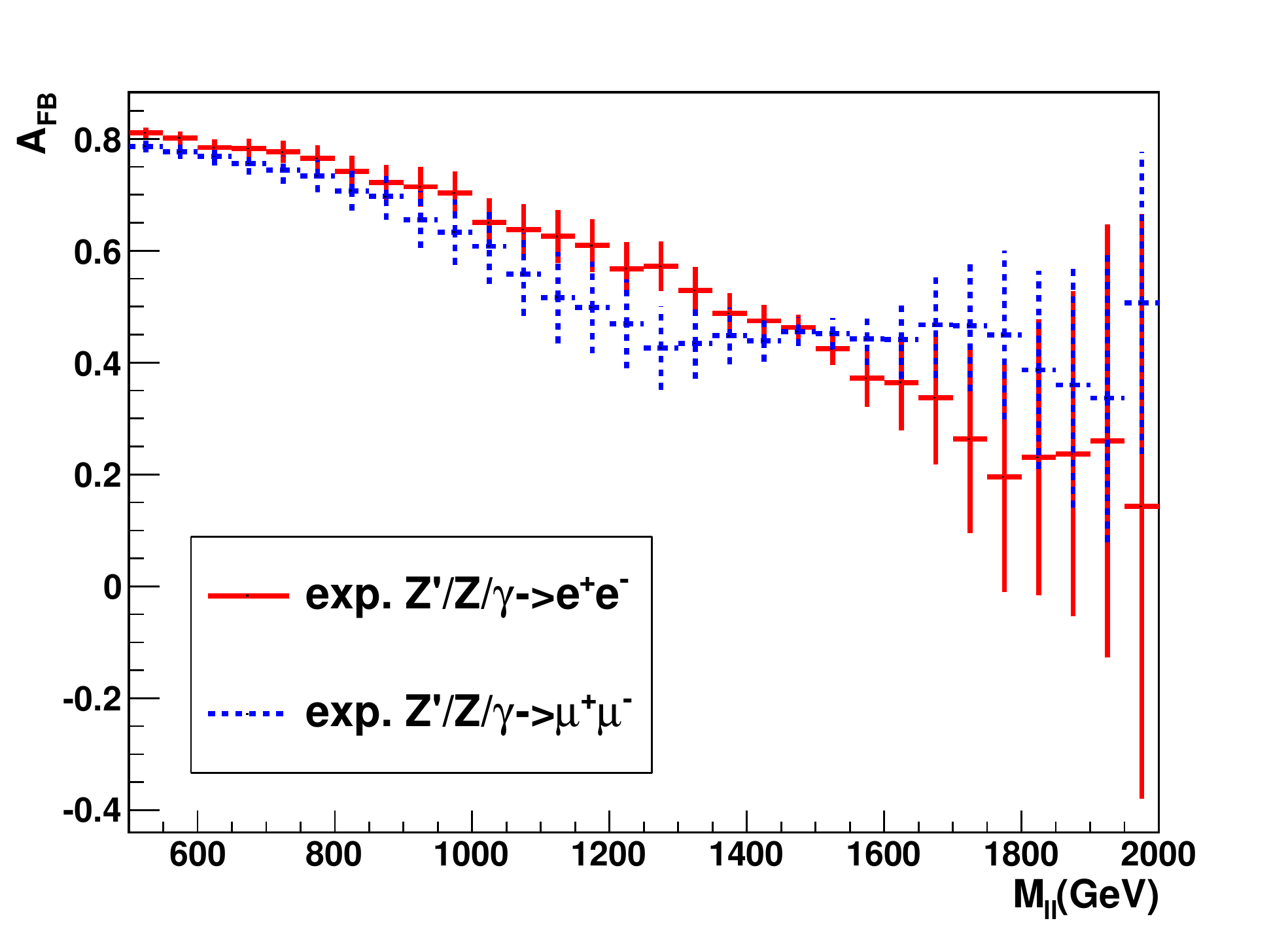}
	\caption{The forward backward asymmetry distribution as a function of the dilepton invariant mass for $M_{Z^{\prime}} = 1.5$ TeV in the case of $\sqrt{s} = 14$ TeV with $500$ fb$^{-1}$ of data. The red lines represent the electron channel and the dotted blue lines denote the muon channel. }
	\label{fig:AFBvsInvM}
\end{figure}

\begin{figure}[tbh!]
\includegraphics[scale = 0.8, angle = 90, width = 80mm, height = 50mm]{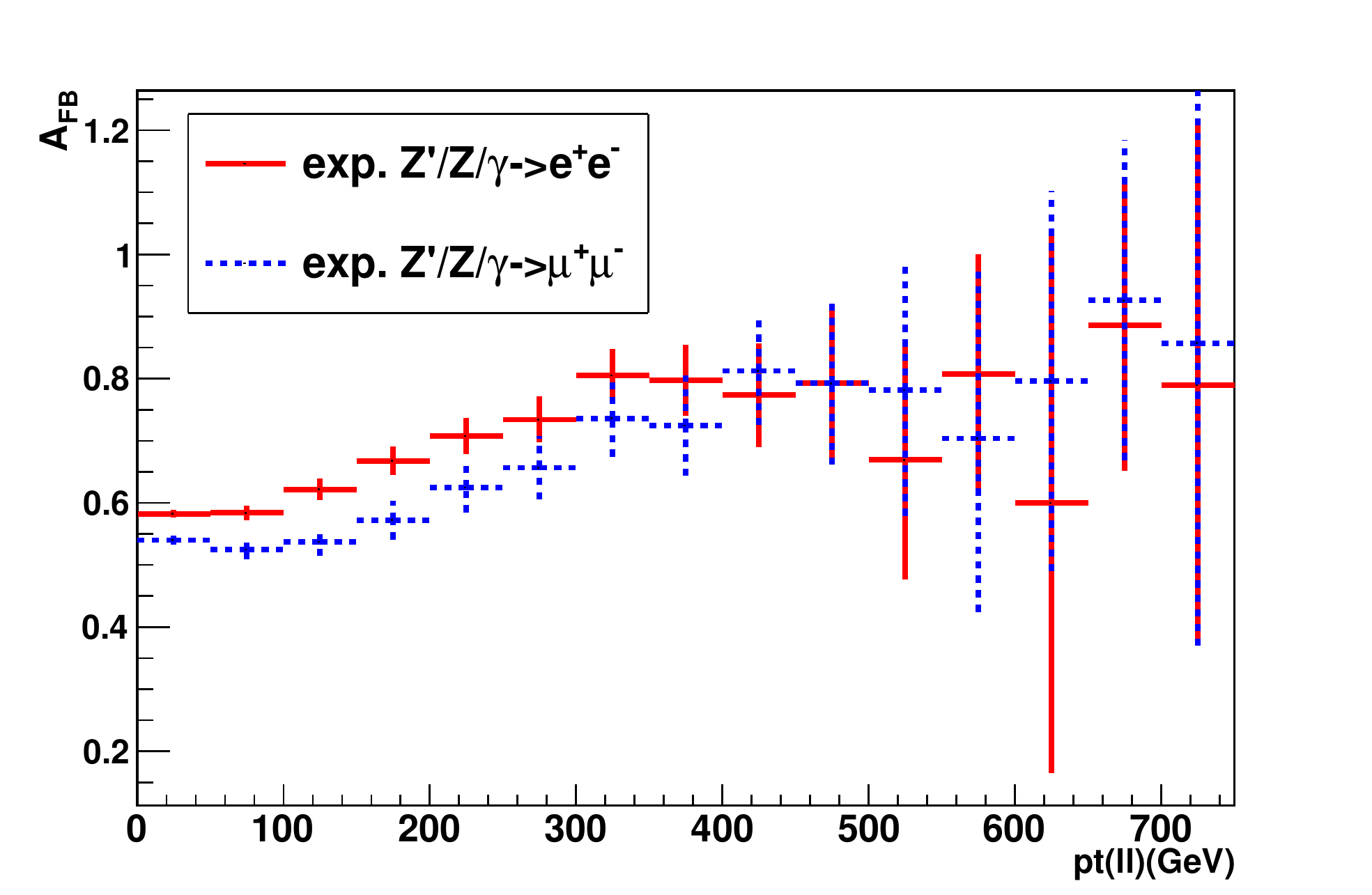}
	\caption{The forward backward asymmetry distribution as a function of the dilepton transverse momentum for $M_{Z^{\prime}} = 1$ TeV in the case of $\sqrt{s} = 10$ TeV with $500$ fb$^{-1}$ of data. The red lines represent the electron channel and the dotted blue lines denote the muon channel. }
	\label{fig:AFBvsPt10TeV}
\end{figure}

\begin{figure}[tbh!]
\includegraphics[scale = 0.8, angle = 90, width = 80mm, height = 50mm]{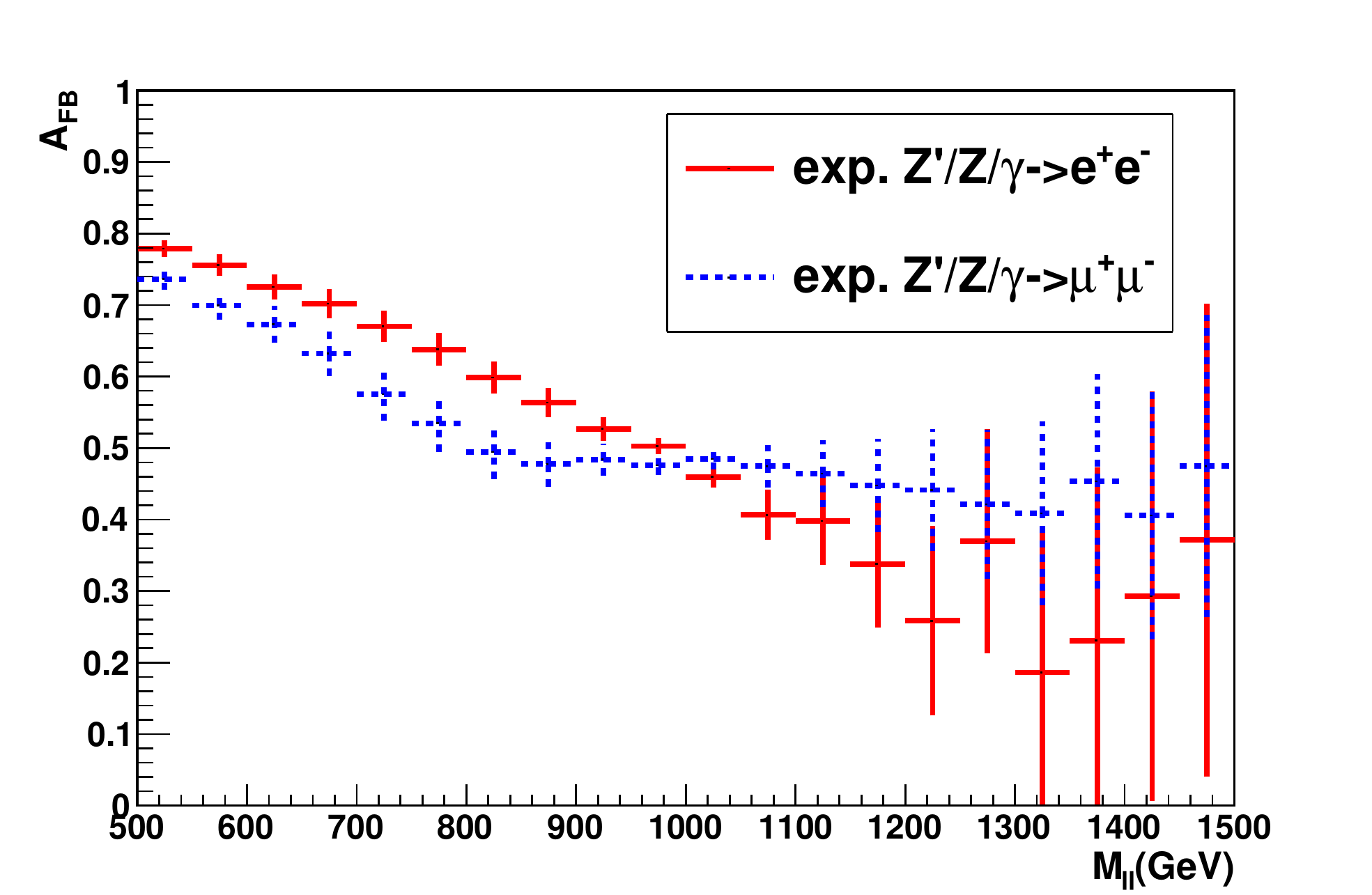}
	\caption{The forward backward asymmetry distribution as a function of the dilepton invariant mass for $M_{Z^{\prime}} = 1$ TeV in the case of $\sqrt{s} = 10$ TeV with $500$ fb$^{-1}$ of data. The red lines represent the electron channel and the dotted blue lines denote the muon channel. }
	\label{fig:AFBvsInvM10TeV}
\end{figure}

In the case of $\sqrt{s} = 14$ TeV with 500 fb$^{-1}$ of data, Fig.~\ref{fig:AFBvsPt} shows the forward backward asymmetry distribution as a function of  the dilepton transverse momentum $p_{t}$ for $M_{Z^{\prime}} = 1.5$ TeV, while Fig.~\ref{fig:AFBvsInvM} shows the forward backward asymmetry as a function of the dilepton invariant mass $M_{\ell\ell}$. The uncertainties correspond to the statistical errors normalized to $500$ fb$^{-1}$ of data. In the case of $\sqrt{s} = 10$ TeV with 500 fb$^{-1}$ of data, Fig.~\ref{fig:AFBvsPt10TeV} and \ref{fig:AFBvsInvM10TeV} show the forward backward asymmetry as a function of the transverse momentum and invariant mass, respectively, for $Z^{\prime}$ mass of 1 TeV. By using the forward backward asymmetry distributions, a clear distinction between the dielectron and dimuon channels can be obtained in the low invariant mass and low transverse momentum regions. As expected, the asymmetry distribution measurements require a large amount of data and thus can only be achieved in the later stages of the LHC operation. 


\section{\label{subSec:bound}Allowed Region of Parameters $a$ and $b$}

\begin{figure}[tbh!]
\includegraphics[scale = 0.8, angle = 0, width = 80mm, height = 80mm]{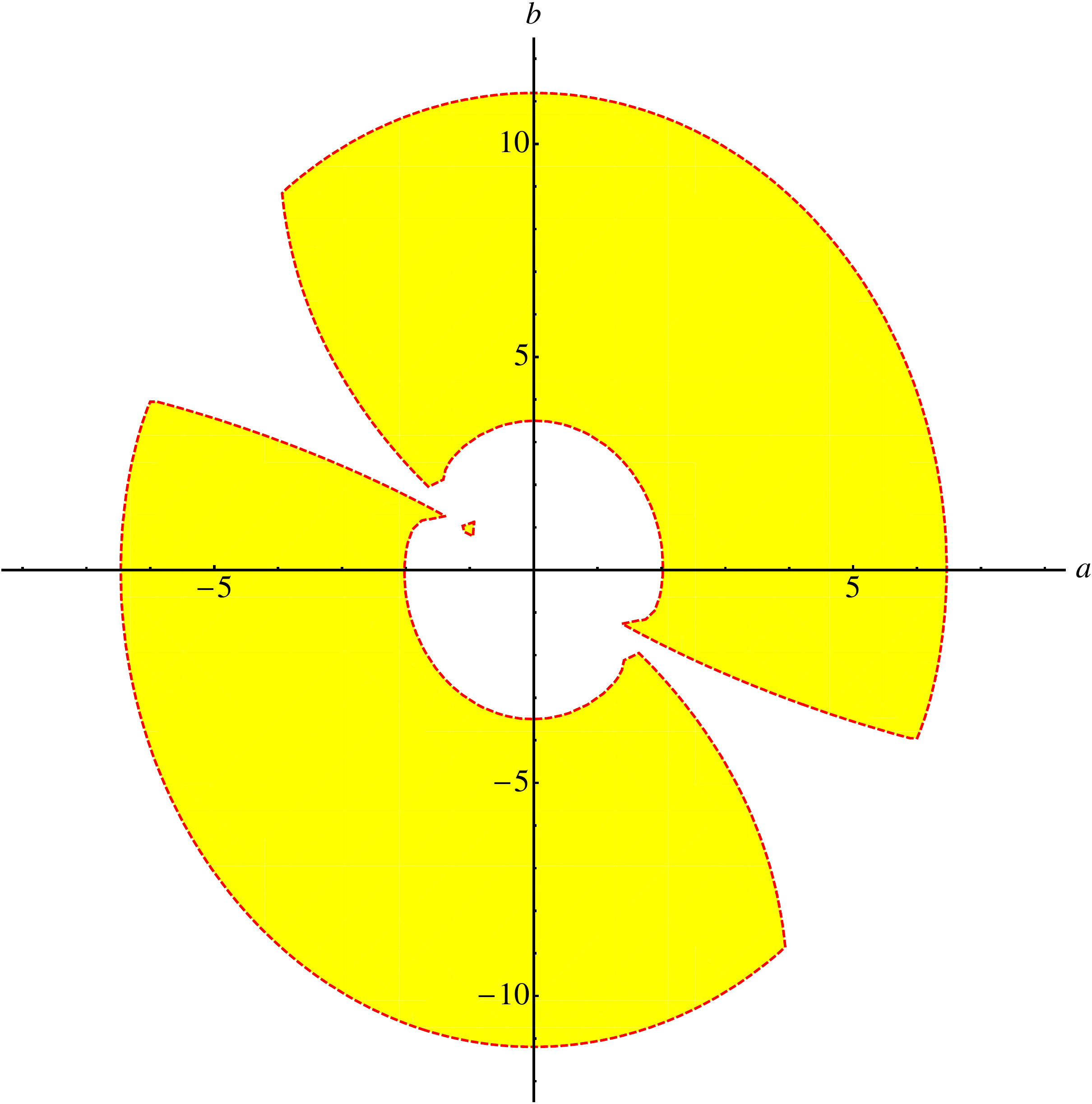}
	\caption{Region of parameter space for (a, b) which satisfies $0.01 < \frac{\Gamma_{Z^{\prime}}}{M_{Z^{\prime}}} < 0.1$ while satisfying other exsiting experimental constraints.}
	\label{fig:bound}
\end{figure}

In the above analyses, we concentrate on only one bench mark point in the parameter space, which gives rise to realistic neutrino mass and mixing patterns.  Without imposing the constraints from the neutrino sector, it is possible to have a more general generation dependent, non-amolaous $U(1)^{\prime}$ model. Fig.~\ref{fig:bound} shows the parameter space of $a$ and $b$ for which the $Z^{\prime}$ discovery is possible. Specifically, the values of $a$ and $b$ in the allowed region gives, $0.01 < \frac{\Gamma_{Z^{\prime}}}{M_{Z^{\prime}}} < 0.1$. We require the $Z^{\prime}$ decay width to be larger than $0.01 M_{Z^{\prime}}$, so that the decay width is wide enough compared to the detector resolution, and thus enabling the $Z^{\prime}$ resonance to be observable. The upper limit of 0.1 for the ratio of the decay width to $M_{Z^{\prime}}$ is imposed to  ensure the narrow width approximation is valid in our analysis. The direct search limits from the Tevatron can be translated into constraints on parameters $C_{u}$ and $C_{d}$ (for the definition of $C_{u,d}$, see Ref.~\cite{ref:CDFz'Con1}.) In the class of non-anomalous models we consider, $z_{u} = z_{d}$, and thus the Tevatron constraints for $M_{Z^{\prime}}$ = 900 GeV~\cite{ref:CDFbound}
\begin{equation}
C_{q} = [(g_{q}^{L})^2 + (g_{q}^{R})^2]B(Z^{\prime} \rightarrow \ell^{+}\ell^{-}) \leq 0.01
\end{equation}
translate into $C_{u} = C_{d} < 0.01$, which are equivalent to. 
\begin{equation}
\frac{\alpha^2(\alpha - 2a)^2}{(8\alpha^2 + 9a^2 + 3b^2)} < 4.5 \; , \quad 
\frac{\alpha^2(\alpha + a)^2}{8\alpha^2 + 9a^2 + 3b^2} < 4.5 \; ,
\end{equation}
for the dielectron and dimuon channels respectively in our model, with $\alpha \equiv -z_{\phi}^{-1} = \frac{a^2 + ab + b^2}{3(a + b)}$. Furthermore, to satisfy the electroweak precision constraints, the parameters (a, b) must be chosen such that $z_{q}^2 < 10^{4}$. Generally, any TeV scale generation dependent $U(1)^{\prime}$ model suffer severe constraints from flavor changing neutral currents. This is easily avoided by choosing $3a$ to be a non-integer, which naturally gives the texture zeros in the (12), (13), (21) and  (31) entries of the charged lepton mass matrix given by Eq.~\ref{eqn:matx}. The FCNC constraints can also be avoided by having a large integral value for $3a$, so that the off diagonal elements in the charged lepton mass matrix are highly suppressed. A large parameter space remains even after removing points for which $3a$ is an integer.

\section{\label{sec:conclusion} Conclusion}

We investigate the collider signatures of a TeV scale non-anomalous $U(1)_{\nu}$ model, which generates at the TeV scale small neutrino masses and their mixing angles. Since the $U(1)_{\nu}$ symmetry, which is different from $U(1)_{\mbox{\tiny B-L}}$, is generation dependent, all gauge anomalies are cancelled with no exotic fields other than the three right-handed neutrinos. Specifically, we have investigated the LHC's discovery potential of such a flavorful $Z^{\prime}$. Through the excess in the dilepton invariant mass distribution,  at the center of mass energy of $\sqrt{s} = 14$ TeV,  the $Z^{\prime}$ of mass up to 4.5 TeV can be discovered at $5\sigma$ with 100 fb$^{-1}$ of data. To establish the flavorful nature of $Z^{\prime}$ requires a distinction between the $e^{+}e^{-}$ and $\mu^{+}\mu^{-}$ channels. While it requires a much higher integrated luminosity, it is possible to distinguish these two decay channels at sufficient significance level.  For the bench mark point we consider in our analysis, at $\sqrt{s}$ = 14 TeV, a $5\sigma$ distinction between the $e$ and $\mu$ channels can be obtained with 500 fb$^{-1}$ of data, for $M_{Z^{\prime}}$ up to 3 TeV. At $\sqrt{s} = 10$ TeV, 8.32 fb$^{-1}$ of integrated luminosity is required for $M_{Z^{\prime}}$ = 1 TeV.  We have also studied the possibility of measuring the $U(1)_{\nu}$ charges of the leptons using the forward backward asymmetry distributions of the dilepton channels. With a sufficient integrated luminosity, it is possible to observe the different asymmetry distributions for the dieletron and dimuon channels in the low invariant mass and low transverse momentum regions. This thus allows to establish the generation-dependent nature of the $U(1)_{\nu}$ model. 

\begin{acknowledgments}
We thank Thomas G. Rizzo for the useful discussion. The work of M-CC was supported, in part, by the National Science Foundation under grant no. PHY-0709742. 
\end{acknowledgments}

\end{document}